\documentclass[12pt]{article}%
\usepackage{amsmath}
\usepackage{amsfonts}
\usepackage{amssymb}
\usepackage{graphicx}%
\begin{document}

\begin{center}
\bigskip

Time Travel Paradoxes, Path Integrals, and the Many Worlds Interpretation

of Quantum Mechanics

\bigskip

Allen Everett

Department of Physics and Astronomy and Institute of Cosmology

Tufts University

\bigskip

\bigskip Abstract

\bigskip
\end{center}

We consider two approaches to evading paradoxes in quantum mechanics with
closed timelike curves (CTCs). In a model similar to Politzer's, assuming pure
states and using path integrals, we show that the problems of paradoxes and of
unitarity violation are related; preserving unitarity avoids paradoxes by
modifying the time evolution so that improbable events become certain. Deutsch
has argued, using the density matrix, that paradoxes do not occur in the
\textquotedblleft many worlds interpretation\textquotedblleft. We find that in
this approach account must be taken of the resolution time of the device that
detects objects emerging from a wormhole or other time machine. When this is
done one finds that this approach is viable only if macroscopic objects
traversing a \ wormhole interact with it so strongly that they are boken into
microscopic fragments. .\pagebreak\qquad

\bigskip

\begin{center}
I. Introduction
\end{center}

There has recently been a good deal of interest in possible spacetimes
containing closed timelike curves (CTCs) arising either from the presence of
traversable wormholes [1] or from the warping of spacetime in such a way as to
allow superluminal travel[2], with the possibility of CTC's as a
consequence[3-5]. A variety of theoretical considerations (e. g., Refs. 6-8),
either general or addressed to specific models, have been advanced which
suggest that the formation of CTC's is not possible. However, while some of
these considerations are very persuasive, none appear conclusive[9].

\qquad In addition to the problems discussed in the references already cited,
CTC's lead to the well known problems with paradoxes arising from the apparent
possibility of inconsistent causal loops. This phenomenon is illustrated by
the \textquotedblleft grandfather paradox\textquotedblright\ occurring
frequently, in various guises, in science fiction, in which one travels back
in time and murders one's own grandfather, thus preventing one's self from
being born and traveling back in time in the first place.

Satisfactory physical theories must avoid giving rise to such
self-contradictory predictions. One approach to achieving this is to impose
consistency constraints on the allowable initial conditions on spacelike
surfaces prior to the formation of the CTC's, thus abandoning the principle
that initial conditions on such surfaces can be chosen at will. For example,
in the case of the grandfather paradox we might insist that the initial
conditions just before the prospective murder include the presence of a
strategically placed banana peel on which the prospective murderer slips as he
pulls the trigger, thus spoiling his aim. One might refer to this approach as
the \textquotedblleft banana peel mechanism\textquotedblright; it leads to a
theory free of logical contradictions, but requires occurrences that would
seem, \textit{a priori}, to be highly improbable. This violates strong
intuitive feelings. These feelings may simply reflect our lack of experience
with phenomena involving CTC's. Nevertheless, a need to invoke constraints on
the choice of initial conditions would be quite disturbing for many physicists
and contribute to an expectation that CTC's are forbidden.

\qquad The first suggestion in the literature that it might be possible to
avoid such paradoxes was due to Echeveria, Klinkhammer and Thorne [10,
henceforth EKT]. For simplicity, these authors formulate the problem in terms
of billiard balls, thereby avoiding questions of free will. They consider a
situation in which, at t = -$\varepsilon$, where $\varepsilon>0$ but may be
taken arbitrarily small, there is a billiard ball (henceforth generally
denoted by BB), which we take to be at the spatial origin; its trajectory is
such that at t = T its leading edge reaches the point \textbf{r}$_{0}$ +
$\Delta$\textbf{r} and enters a wormhole which connects the spacetime point
(\textbf{r}$_{0}$\textbf{\ + }$\Delta$\textbf{r}, t + T) to (\textbf{r}$_{0}$,
t ). (A procedure for creating such a wormhole is discussed in Ref. 10. We
take $\Delta$r, the spatial distance between the wormhole mouths, to be small
compared with r$_{0}$ , and will in general ignore it; however, $\Delta$r
cannot vanish if the wormhole persists over a time interval T$_{0}$
$>$
T, as one needs to introduce some separation between the two wormhole mouths
if they overlap in t. In general we will assume T$_{0}\ll$ T. We take the
internal length of the wormhole to be small compared to r$_{0}$, and will
often work in the approximation in which the two mouths of the wormhole are
simply identified with one another.) Upon emerging from the wormhole mouth at
t = 0, the BB may interact with its \textquotedblleft younger
self\textquotedblright\ which has not yet entered the wormhole.
(\textquotedblleft Younger\textquotedblright\ here means younger in terms of
the ball's \textquotedblleft personal\textquotedblright\ time, i. e., the
proper time $\tau$ measured on a clock attached to the ball.) An inconsistent
causal loop, analogous to the \textquotedblleft grandfather
paradox\textquotedblright, can then occur as the result of a BB trajectory
such that, on emerging from the wormhole, the ball undergoes a head on
collision with its younger self, deflecting the latter so that it does not
enter the wormhole in the first place. However, as EKT point out, in the
presence of CTC's the trajectory is not unique, and there are also solutions,
with the same initial conditions, which give rise to consistent causal loops;
e. g., a glancing collision may occur which deflects the ball's younger self
so that its trajectory through the wormhole results in the required glancing
collision. EKT then suggest adopting a consistency principle according to
which only self-consistent solutions are to be considered physical. The EKT
consistency principle places constraints on the allowable initial conditions
within the region containing CTC's, but does not constrain the initial
conditions that may be imposed outside that region. This idea seems physically
attractive, and the discussion in Ref. [10] had much to do with stimulating
interest in time travel as a subject possibly deserving of serious study.

It is, however, far from clear that the consistency principle always allows
one to avoid paradoxes. Because of the variety of types of collision, ranging
from glancing to head on, which can occur between two spheres, EKT were able
to find self-consistent solutions, in fact an infinite number of them, for a
wide variety of initial conditions. \ However, it is difficult to see how this
can be true in general. For example, suppose we place at the early time (t =
0) mouth of the wormhole a device to detect the BB if it emerges. (One might,
e. g., have a spherical grid of current carrying wires enclosing the wormhole
mouth at t = 0 thin enough to be broken by the BB with its given speed and
spaced closely enough that a BB cannot emerge from the wormhole without
breaking at least one of the wires). Suppose further that we connect the
detector in such a way that, if a BB is detected, a signal is sent at light
speed activating a mechanically operated shutter, which deflects the incident
ball at some later point on its path so that it does not enter the late-time
wormhole mouth. One can include a reqirement that the signal is sent only if a
BB emerges from the wormhole before the incident ball reaches the shutter.
\ This eliminates the possibility of a second, consistent, solution, in whuch
the shutter starts to close just as the ball passes through it, resulting in a
self-consistent time delay in which the ball emerges from the wormhole and
causes the shuter to close just as the incident ball reaches it. This
arrangement is a modification of somewhat similar ones discussed by Novikov
[11] in which consistent solutions exist. In the case discussed here, however,
it is difficult to see how there can be any self-consistent, and hence
physically acceptable, solution. Thus we seem to be back to the grandfather
paradox in the form of a BB which enters the wormhole if and only if it does
not enter the wormhole.

We will illustrate these ideas below, making use of a quantum mechanical model
due to Politzer [12, henceforth HDP] which is simple enough to be calculable
but has many of the physical features of the BB-wormhole system just
discussed. In this model, systems are treated as being in a pure quantum
state, as in standard quantum mechanics, even in a spacetime region containing
CTC's, and the path integral formalism is used; the treatment in HDP is
limited to calculating amplitudes for the case of initial and final states at
times, respectively, before and after the era containing CTC's.   model the
Hamiltonian, H, can be chosen so that there are self-consistent solutions; in
accordance with the EKT principle, these solutions can be taken to be the only
ones which are physically relevant. We also find, at least in the HDP model,
that when unique consistent solutions exist, their time evolution is governed
by a unitary operator, so that the probability interpretaton of quantum
mechanics can be preserved. Hence, if consistent solutions exist, one has a
quantum mechanical theory which, in the presence of CTC's, differs from
standard quantum mechanics only in the imposition of the EKT criterion for
physically relevant solutions, and in the fact that the uniqueness problem
remains if there is more than one self-consistent solution.

However, it is also possible, as we will see below, to choose a Hamiltonian in
the region of CTC's in the HDP model for which no self-consistent solutions
exist, in contrast with the examples in Refs. 10 and 11. One can attempt to
restore consistency by projecting out only that part of the path integral
expression for the wave function at t
$>$
T which comes from paths satisfying the consistency condition. However, as
seen in HDP, this results, because of the consistency requirement, in a time
evolution of the wave function which is controlled by a nonunitary operator X
$\neq$ exp(-iHt). One may preserve the probability interpretation of the final
state wave function by renormalizing the operator X by a factor which depends
on the initial state and which thus introduces nonlinearity into the time
evolution. Unexpectedly, this has consequences for the time evolution of the
system for t
$<$
0, i. e., for times \textit{before} CTC's occur, as first observed by
Hartle[13]; a more intuitive argument based on the requirement for a
consistent probability interpretation is given in HDP. An alternative
procedure for renormalizing X by matrix multiplication, proposed by
Anderson[14], avoids violations of causality at t
$<$
0, though perhaps at the cost of discarding essential physics.\ \ Both
procedures, in effect, lead to the \textquotedblleft banana
peel\textquotedblright\ mechanism, since one finds that the presence of
potential paradoxes insures the occurrence of \textit{a priori} improbable
events either before or during the era of CTC's. \ 

Hence, in the general case, the HDP approach avoids the problems associated
with inconsistent causal loops only if, in the presence of CTC's, fundamental
axioms of quantum mechanics are abandoned. In particular, the time evolution
operator which transforms the wave function at t$_{1}$ into that at t$_{2}$ is
no longer unitary and is not given by U(t$_{1}$, t$_{2}$) = exp(-iHt), with t
= t$_{1}$ -- t$_{2}$.Moreover, the preservation of a consistent probability
interpretation requires the introduction of rather \textit{ad hoc} procedures,
possibly involving \ violations of causality in the era before CTC's are formed.

It would thus be interesting to find a model-independent approach In which the
existence of CTC's does not lead to inconsistent causal loops; the existence
of such a theory would remove one of the theoretical (or perhaps
psychological) objections to CTC's. Hopefully this would avoid the nonunitary
time evolution operators, and the consequent difficulties with conservation of
probability, which arise in HDP when inconsistent causal loops are present.

Science fiction writers often avoid causal paradoxes in stories involving time
travel by invoking the idea of \textquotedblleft alternate
universes\textquotedblright. At first sight this idea seems devoid of any
physical foundation. However, the many-worlds interpretation (MWI) of quantum
mechanics due to Hugh Everett III[15] does introduce ideas which have some
resemblance to the alternate universes of science fiction; it also provides an
interpretation of quantum mechanics which seems difficult or impossible to
distinguish experimentally from the more conventional one, and which some
might argue is intellectually more satisfying.

It thus seems natural to ask whether the MWI might provide a way out of the
problems of logical consistency raised by CTC's. Deutsch [16, hereafter
referred to as DD] has discussed this question. He argues that inconsistent
causal loops do not occur in the MWI because, loosely speaking, the pairs of
seemingly inconsistent events (e. g., one's birth and one's murdering one's
grandfather) occur in different \textquotedblleft universes\textquotedblright%
\ and hence are not logically contradictory. In Deutsch's approach the MWI
becomes more than a mere interpretation of quantum mechanics; in the presence
of CTC's it has experimental consequences.

The approach in DD actually involves assumptions that go beyond simply
adopting the MWI. The cost of preserving unitarity, or more precisely,
conservation of probability, is that, in the presence of CTC's, a system must
in general be described by a density matrix, not a wave function. As we will
discuss, in the absence of self-consistent solutions, pure states necessarily
evolve into mixed states in the region containing CTC's if the violations of
unitarity in the HDP approach are to be avoided. Thus, in the same situations
in which unitarity fails in the model in HDP, the approach in DD requires one
to formulate the theory in terms of the density matrix. The time evolution
equation of the density matrix in DD is given, as usual, by
\begin{equation}
\varrho(t_{2})=\text{U}^{-1}(t_{2},t_{1})\varrho(t_{1})\text{U}(t_{2},t_{1})
\tag{1}%
\end{equation}
where (in units with $\hbar$ = 1) \ U(t) = exp(-iHt) and H is the Hamiltonian;
this ensures the preservation of the probabilistic interpretation of $\varrho
$. Moreover, Eq. (1) is taken to be valid at all values of t so that the
theory determines $\varrho$ during the era in which CTC's exist, as well as
before and after. However, the concepts of \textquotedblleft mixed
state\textquotedblright\ and \textquotedblleft density
matrix\textquotedblright\ in DD are different than in conventional quantum
mechanics, where \textquotedblleft mixed state\textquotedblright\ refers to an
ensemble of identically prepared systems whose statistical properties are
given by the density matrix. In DD the term \textquotedblleft mixed
state\textquotedblright\ refers to a single system not in a definite quantum
state and described not by a wave function but a density matrix. The diagonal
elements of $\varrho$ in, say, the R-representation, where R is an observable,
give the probabilities of observing the possible outcomes of a measurement of
R on that single system, and $\varrho$ will not, in general, satisfy the
condition $\varrho^{2}=\varrho$ characteristic of a pure state.

Working only with density matrices and mixed states of the type just discussed
goes beyond, at least in principle, simply adopting the MWI as presented in
Ref. 14 which deals with systems in pure states described by wave functions.
In the MWI, suppose we begin with an object whose initial wave function $\psi$
= c$_{1}$u$_{1}$(R$_{j}$) + c$_{2}$u$_{2}$(R$_{j}$) where R$_{j}$ are the
eigenvalues of an observable R describing the object, and the u$_{i}$ are
eigenstates of R with eigenvalues R$_{1}$ and R$_{2}$. . Let the value of R be
measured by a macroscopic measuring apparatus which is left in a state with
wave function $\phi_{i}$(q$_{k}$) when the measurement yields the result
R$_{i}$ , where the $\phi_{i}$ are eigenstates of an observable Q with
eigenvalues q$_{k}$ giving the internal state of the measuring apparatus.
According to the MWI, the system of object plus apparatus will be described
after the measurement by a wave function, f(R$_{j}$,q$_{k}),$where \ \ \
\begin{equation}
f(R_{j},q_{k})=c_{1}^{\prime}\phi_{1}(q_{k})u_{1}(R_{j})+c_{2}^{\prime}%
\phi_{2}(q_{k})u_{2}(R_{j}) \tag{2}%
\end{equation}
\linebreak and
$\vert$%
c$_{i}^{\prime}$%
$\vert$
=
$\vert$%
c$_{i}$%
$\vert$%
. Hence the object-apparatus system remains in a pure state. However, because
of the complexity of the internal structure of the measuring apparatus, the
eigenvalues q$_{i}$ are highly degenerate. Hence the two terms on the right
side of Eq. (2) actually represent effectively infinite sums of terms with
varying phases. Thus, once the measurement interaction is over, the two terms
on the right side of Eq. (2) become decoherent and matrix elements of
operators between states with different $\phi_{i}$ effectively vanish. This is
the reason the \textquotedblleft worlds\textquotedblright\ in which Q has
different well-defined values are unaware of one another so that the MWI, at
least in the absence ofCTC's, is without observable consequences.

From the foregoing discussion we see that the approach in DD to resolving the
paradoxes associated with time travel involves modifying fundamental
principles of quantum mechanics; it certainly goes beyond simply adopting the
MWI. We will refer to this approach from now on as the \textquotedblleft mixed
state MWI\textquotedblright\ (MSMWI) to distinguish it from the original many
worlds interpretation of Ref. 14. However, despite the differences in
principle, in practice, when dealing with macroscopic systems, the
\textquotedblleft mixed states\textquotedblright\ which occur in Deutsch's
approach are very similar to the nearly decoherent states which occur in the
MWI following a measurement, so one might feel that the departure from
standard quantum mechanics is relatively minor, and perhaps plausible.

However, as we argue below, once an observation has been made as to whether a
BB has or has not emerged from the wormhole the states corresponding to these
two possibilities become decoupled, just as in the case of the different
\textquotedblleft worlds\textquotedblright\ of the MWI when no CTC's are
present. As a result, in situations where, classically, there would be an
inconsistent causal loop, while the front part of an object traveling backward
in time emerges from the wormhole in a different "world"; another part emerges
in the same \textquotedblleft world\textquotedblright\ which contains its
younger self, contrary to the proposal in DD. As a result, in the case of
macroscopic objects, when proper account is taken of the finite time required
for the object to emerge from the wormhole and be detected, one finds that no
self-consistent solutions in which the object passes intact through the
wormhole exist in the MSMWI. The object is sliced into two, or more generally
into many, pieces, in passing through the wormhole, with different pieces
winding up in different \textquotedblleft worlds\textquotedblright, i. e., in
states of the system labeled by different readings of a macroscopic measuring
device. Thus, in the MSMWI, wormholes (or other time machines) which can be
traversed intact by macroscopic objects cannot exist. If the MSMWI is correct,
such objects must necessarily undergo violent interactions with the time
machine which cause the object to disintegrate.

The organization of the remainder of the paper is as follows. In Section II we
discuss the quantum mechanical formulation of the consistency condition in the
presence of CTC's in terms of its implications for the time evolution operator
of the wave function. In Section III, we consider the model in HDP in cass
where the operator U does, and does not, give rise to the existence of
consistent solutions, and observe the connection in the model between the
existence and uniqueness of consistent solutions and unitarity. In Section IV
we review in detail the density matrix approach in DD, and its connection to
the MWI, and discuss the relation between the absence of consistent solutions
and the transformation of an initial pure quantum state into a mixed state in
the region containing CTC's. In Section V we examine, following DD, how the
MSMWI might resolve the analog of the \textquotedblleft grandfather
paradox\textquotedblright\ in the case of a microscopic object, such as an
electron, traveling backward in time. In Section VI we analyze in detail the
difficulties which arise when one attempts to extend the MSMWI to macroscopic
objects. We conclude briefly in Section VII.

\begin{center}
II. The Consistency Condition for Wave Functions
\end{center}

Here we assume that the rules of quantum mechanics are unchanged in the
presence ofCTC's except for the imposition of a consistency requirement, whose
formulation we wish to examine. Suppose that at t = $-\varepsilon$, where
$\varepsilon$ is infinitesimal, we have an incident BB at the origin whose
trajectory is such that its leading edge reaches the wormhole at t = T. We
take the BB's proper time $\tau$ to be the position of the hand of a clock
attached to the ball, which we can treat as a dynamical observable. In
contrast, t, the evolution parameter for wave functions, may be thought of as
the common reading of a network of synchronized clocks remaining at rest
relative to one another. We suppress the (many) other internal variables in
addition to $\tau$ associated with the internal structure of the BB. For t
$<$
T, $\tau$ = t; however, neglecting the travel time through the wormhole,
$\tau$ = t + T for $\tau$
$>$
T, i. e., for the ball which emerges from the wormhole. Although a classical
object, we assume the BB is, in quantum mechanics, described at the
fundamental level by a wave function $\psi_{1}$(\textbf{r} , $\tau$, t) whose
dependence on the dynamical variables \textbf{r }and $\tau$ at t = -$\epsilon$
is peaked about their classical values (r = 0 and $\tau$ = -$\epsilon$ ) with
negligible spread; by continuity this should also be true of $\psi_{1}%
$(\textbf{r}, $\tau$, +$\epsilon$). However, there may now be what appears to
be a second ball emerging from the wormhole. Since we expect the wave function
of the ball near the origin to be determined by continuity, we take the most
general form of the wave function for the system to be%

\begin{equation}
\bigskip\qquad\psi(\mathbf{r},\tau,\mathbf{r}\prime,\tau\prime,n\prime
,t=\epsilon)=\psi_{1}(\mathbf{r},\tau,\epsilon)\psi_{2}(\mathbf{r}\prime
,\tau\prime,n\prime,\epsilon) \tag{3}%
\end{equation}
\linebreak where the variable n' is an occupation number with two possible
values, 1 and 0, denoting, respectively, the presence or absence of a BB
emerging from the wormhole. [Thus, e. g., if the incident ball always goes
through the wormhole $\psi_{2}$(n' = 0) = 0. Excluding the possibility n'
$>$
1 corresponds to the assumption that the incident BB, if it emerges from the
wormhole mouth at t = 0, is directed in such a way that it does not reenter
the wormhole mouth at t = T.] For n' = 1, \textbf{r}' and $\tau$' are position
variables for the emerging BB and the hand of its clock, so that $\psi_{2}$(n'
= 1, t = $\epsilon$) is peaked around the values \textbf{r}' = \textbf{r}%
$_{0}$ and $\tau$' = T + t$_{0}$ + $\epsilon$, where t$_{0}$ is the transit
time of the BB through the wormhole; since $\epsilon$ is arbitrarily small,
and the internal length of the wormhole is taken to be such that t$_{0}$
$<$%
$<$
T, , $\tau$' $\approx$ T. Note that the product form of the wave function, Eq.
(3), which we obtained by continuity in time from the initial condition at t =
-$\epsilon$, also follows from the reasonable assumption that the two balls
(actually the younger and older versions of the same ball) will not yet have
interacted at t =+ $\epsilon$.

While traversing the wormhole, we take the subsystem in the vicinity of
\textbf{r}$_{o}$ to be \ isolated. For this subsystem, $\tau$, though it may
be regarded as a dynamical observable by outside observers, plays the role of
the time evolution parameter. The BB evolves through the wormhole in the
direction of increasing $\tau$ or decreasing t with the evolution governed by
a Hamiltonian H', the Hamiltonian of the isolated subsystem. (Instead of a
billiard ball, one can picture this in terms of an isolated spaceship inside a
superluminal Alcubierre warp bubble.[5] Passengers on the ship would see their
world governed by a quantum mechanics in which the reading of clocks on the
spaceship would play the role of the time evolution parameter even though the
hands of the clocks appear to run backwards to outside observers in some
Lorentz frames.)

From the foregoing, we conclude that the wave function at t = T can be
obtained from that at t = $\epsilon$ by an operator of the form
\begin{equation}
U(T)=u(T)\otimes exp[iH\prime to) \tag{4}%
\end{equation}
\linebreak where u(T) is an operator which acts on $\psi_{1}$(\textbf{r},
$\tau,\epsilon$, ) and leaves $\psi_{2}$( \textbf{r}', $\tau$', n', $\epsilon
$) invariant since its evolution is governed solely by exp(iH't$_{0}$). The
sign of the exponent is due to the fact that $\tau$ decreases in going from
the t = 0 to the t = T mouth of the wormhole. In the approximation that the
wormhole mouths are identified so that t$_{o}$ = 0,%

\begin{equation}
U(T)=u(T)\otimes I \tag{4'}%
\end{equation}
and the wave function of the system at t = T has the form

\qquad%
\begin{equation}
\psi(\mathbf{r},\tau,n,\mathbf{r}^{\prime},\tau\prime,n\prime,T)=\psi
_{1}(\mathbf{r},,n,T)\psi_{2}(\mathbf{r}\prime,\tau\prime,n\prime,\epsilon)
\tag{5}%
\end{equation}
\linebreak where, at t = T, we include an occupation number variable n for the
part of the system separated from the wormhole. (At t = 0, n = 1 from the
initial conditions.) The direct product structure of (4) or (4') and the
product form of (5) are consequences of the fact that H' depends only on the
coordinates of the subsystem traversing the wormhole. Thus H' cannot generate
any correlations at t = T between the BB's with coordinates \textbf{r} and
\textbf{r}'. If t$_{0}\neq$ 0, a factor of exp(-iE$_{n^{\prime}}t_{0}$) must
be included on the right of Eq.(5), where E$_{0}$ = 0 and E$_{1}$ = E$_{BB}$,
the energy associated with the presence of a BB; in general, this factor, even
if present, does not affect the subsequent discussion

Note, from Eqs. (3) and (5), that the wave function itself is not continuous
across the wormhole; only the factor $\psi_{2}$ is continuous as a consequence
of Eq. (4'). If one begins with a system in a pure state (and thus describable
by a wave function), then since it is being evolved by a unitary operator the
overall system remains in a pure state. The wave function at t = $\epsilon$
has the product structure of Eq. (3) so that each subsystem is separately in a
pure state. The continuity across the wormhole, expressed by Eq. (4), then
guarantees that structure is preserved in Eq. (5) even though the subsystems
outside the wormhole interact with one another between t = 0 and t = T; the
subsystem at the wormhole mouth thus remains separately in a pure state. Thus
if $\psi_{1}$(T) and/or $\psi_{2}$(T) are superpositions of states with
different occupation numbers, the values of n and n' must be uncorrelated.

\begin{center}
III.CTC's and the Quantum Mechanics of Pure States
\end{center}

We begin this section by summarizing Politzer's model referred to above. The
model drastically truncates spacetime to a space with two fixed points with
coordinates z =z$_{1}$and z = z$_{2}$. For z = z$_{1}$, -$\infty$
$<$
t
$<$
$\infty$. However, at z = z$_{2}$ there is a time machine in the form of a
wormhole connecting t = 0 and t = T; we neglect the transit time through the
wormhole, t$_{0}$, so that the wormhole simply identifies the spacetime points
(z$_{2}$, 0) and (z$_{2}$, T). A particle at z = z$_{2}$, t
$<$
0 is taken to enter the wormhole at t = 0 and emerge at t = T to move on into
the future, while a particle at z$_{2}$ in the range 0
$<$
t
$<$
T enters the t = T mouth of the wormhole and emerges at t= 0, following a
worldline which is an endless CTC at constant z = z$_{2}$. The physical system
is taken to be a single fermion field. Hence the occupation numbers n and n'
are restricted to 0 or 1. This models the situation discussed in Section II,
where n = 1 is an initial condition at t = 0, and n'$\leq$ 1 as a result of
the assumed trajectory of a BB emerging from the wormhole. \ The states of the
system lie in a Hilbert space a basis for which is provided by the four states
$\uparrow_{1}\uparrow_{2},\uparrow_{1}\downarrow_{2}$, $\uparrow_{2}%
\downarrow_{1}$, and $\downarrow_{1}\downarrow_{2}$, where $\uparrow
_{1}\uparrow_{2}$, e. g., corresponds to occupation number 1 for the fermion
states at both z$_{1}$ and z$_{2}$-. Henceforth we will omit the subscripts 1
and 2, unless needed for clarity, and simply adopt the convention that the
first and second arrows in a pair denote, respectively, the occupation numbers
at z$_{1}$ and z$_{2}$. The field is taken to have effectively infinite mass
so that kinetic energy terms in the energy can be ignored. The notation
$\uparrow$ and $\downarrow$ for the occupied and unoccupied states,
respectively, is motivated by the fact that the model is mathematically
equivalent to the presence of spin-1/2 particles, each with possible spin
states up and down, at z$_{1}$ and z$_{2}$. We will always deal with the case
where the state at z$_{2}$ is unoccupied except for 0
$<$
t
$<$
T. During this interval the system is governed by a Hamiltonian H whose
general form is that of an arbitrary 4x4 Hermitian matrix whose matrix
elements are constants, since the particle positions are takem to be given by
the occupation numbers and there is no kinetic energy. The assumed freedom to
choose the Hamiltonian governing the time evolution in the interval 0
$<$
t
$<$
T is a natural representation in the model of the assumed freedom to impose
arbitrary initial conditions at t = 0 before CTC's exist.

Let $\Psi$(t) be the state vector of the system, which, in the absence of
CTC's, would obey the equation

\qquad\qquad\qquad\qquad%
\begin{equation}
\Psi(t)=\exp(-iHt)\Psi(+\epsilon) \tag{6}%
\end{equation}
\linebreak Note that an arbitrary 4x4 unitary matrix U(T) can be written as
exp(-iHT) where the Hermitian operator H is a generator of the group U(4);
hence the freedom to choose H arbitrarily translates into the freedom to
choose an arbitrary U(T).

We will take the model in HDP to give a calculable qualitative guide to the
behavior of our system of the billiard ball that travels back in time and
interacts with itself. We let z$_{1}$ be the position of the incident BB at t
= -$\epsilon$ and z$_{2}$ the position of the wormhole. As in the HDP model,
the state of the system at t = +$\epsilon$ lies in the same Hilbert space as
before, where now, e. g., the state $\Psi$(+$\epsilon$) = $\uparrow\uparrow
$corresponds to the state with both an incident BB at z$_{1}$ and a ball at
z$_{2}$ which emerged from the wormhole. The behavior of the BB system is, of
course, more complicated than that of the model. The actual trajectory of the
balls is not one of constant z; the incident ball may move from z$_{1}$ to
enter the wormhole at z$_{2}$, while a ball emerging from the wormhole at t =
0 might be directed along a trajectory reaching z = z$_{final}$at t = T,
where, for simplicity, we could take z$_{final}$= z$_{1}$. Moreover during the
interval 0
$<$
t
$<$
T the two BBs may interact with each other in complicated ways. For example:
they may collide;or they may hit switches which cause shutters to be closed,
diverting or stopping one or both of the initial balls..However, the state of
the system at t = T - $\epsilon$ (or at least that part of it in which we are
interested), can again be described in terms of the same Hilbert space, where
now $\uparrow\uparrow$ is the state in which balls are present at both z$_{2}%
$, on the verge of entering the t = T mouth of the wormhole, and at
z$_{final}$. Hence, for a prescribed z$_{final}$, which we will take to be
z$_{1}$, the range of possible time evolutions of the billiard ball system is
given by the range of possible 4x4 unitary matrices U(t), as in HDP. \ Since
the HDP\ model also enforces consistency between the two mouths of the
wormhole, it contains much of the essential physics of the BB-wormhole system.
Hence it seems reasonable to hope that HDP provides a correct qualitative
descripton of the behavior of the latter system.

We are interested in the case where the initial state at t = -$\epsilon$ is
$\uparrow\downarrow$; i e., we have an initial particle at z$_{1}$ but not at
z$_{2}$. (We will always impose the initial condition at z = z$_{2}$ that \ no
BB is present, i. e. that n' = 0, at t = -$\epsilon$.) We assume, from
continuity, that the occupation number at z$_{1}$ at t = +$\epsilon$ remains
equal to 1, so that the two possible states of the system at that time are
$\uparrow\uparrow$ and $\uparrow\downarrow$. Politzer in HDP obtains the
amplitude for finding occupation number i, (i = 0 or 1) at z$_{1}$ at t = T +
$\epsilon$, given occupation number 1 at t = -$\epsilon$, by using the Feynman
path integral to sum over paths, subject to the consistency condition that the
occupation number at z = z$_{2}$ be the same at t = 0 and t = T since these
two points are identified. The result, as given in HDP, is that the amplitude
is given by X$_{i1}$, where \ 

\qquad\qquad\qquad%
\begin{equation}
X_{ij}=\sum_{k}<ik|U(T)|jk> \tag{7}%
\end{equation}
\linebreak where
$\vert$%
ik%
$>$
denotes a state with occupation numbers n = i and n' =k at z$_{1}$ and z$_{2}%
$, respectively. Note that X$_{ij}$ involves a the trace of the 2x2 matrix
U$_{ik;jn}$ in the occupation number space at z$_{2}$; \ it does not,
therefore, depend on the choice of the set of two orthonormal basis states in
that space.

We define
$\vert$%
X(j)%
$\vert$%
$^{2}$ = $\sum_{i=0}^{1}$%
$\vert$%
X$_{ij}$%
$\vert$%
$^{2}$., j = 0, 1. It is pointed out in HDP that the overall normalization of
\textbf{X} may be multiplied by a state independent factor that can be
absorbed in the functional measure. However, unitarity requires%

\begin{equation}
|X(1)|^{2}=|X(0)|^{2} \tag{8}%
\end{equation}

We consider below several different cases, differing in the form of U in Eq.
(7). We will always assume, on physical grounds, that U is such that
\begin{equation}
U(T)\downarrow\downarrow=\downarrow\downarrow\tag{9}%
\end{equation}
i. e., we assume that if no particle is present either at z$_{1}$ or z$_{2}$
at t= +$\epsilon$, none will be present at t = T -- $\epsilon$.

\begin{center}
Case 1
\end{center}

First take the example in which the incident BB is on a trajectory to carry it
into the wormhole mouth at t = T, and a ball emerging from the wormhole mouth
at t = 0 is directed onto a trajectory taking it to z = z$_{1}$ at t = T. Then%

\begin{equation}
U(T)\uparrow\uparrow=\uparrow\uparrow\tag{10}%
\end{equation}
\qquad\qquad\qquad\qquad and%

\begin{equation}
U(T)\uparrow\downarrow=\downarrow\uparrow\tag{11}%
\end{equation}
\linebreak Hence the matrix elements
$<$%
ik%
$\vert$%
U%
$\vert$%
1k%
$>$
appearing on the right side of Eq. (7) are nonvanishing only for i = k = 1.
For the U(t) in Eqs. (10) and (11), the states $\uparrow\uparrow$ and
$\uparrow\downarrow$ at t = +$\epsilon$ thus give, respectively, completely
consistent and completely inconsistent solutions. For the completely
inconsistent solution,$\uparrow\downarrow$, the wave functions at z = z$_{2}$
at t = 0 and t = T are orthogonal and have no overlap; the incident BB enters
the wormhole at t = T, but no BB emerges at t = 0. For the consistent
solution,$\uparrow\uparrow$, the wave function at t = T -- $\epsilon$ has the
product form given in Eq. (5), as demanded by continuity.

It follows from Eq. (7) that, for U given by Eqs. (10) and (11), X$_{01}$ = 0
because
$<$%
0k%
$\vert$%
U(T)%
$\vert$%
1k%
$>$
= 0. Hence (7), (10), and (11) give%

\begin{equation}
|X(1)|^{2}=|X_{11}|^{2}=1 \tag{12}%
\end{equation}
\qquad\qquad\qquad\qquad\linebreak Eqs. (9)-(11), coupled with unitarity,
imply that

\qquad\qquad\qquad\qquad%
\begin{equation}
U(T)\downarrow\uparrow=\uparrow\downarrow\tag{13}%
\end{equation}
\qquad\qquad\qquad\qquad\qquad\linebreak so that one sees, from (9) and (12),
that there is also one consistent and one inconsistent solution for the case
that there is no incident BB at t = -$\epsilon$. As a result the calculation
of
$\vert$%
X(0)%
$\vert$%
$^{2}$ exactly parallels that for
$\vert$%
X(1)%
$\vert$%
$^{2}$ and we find

\qquad\qquad\qquad%
\begin{equation}
|X(0)|^{2}=|X_{00}|^{2}=1 \tag{14}%
\end{equation}
\qquad\qquad\qquad\linebreak\linebreak From Eqs. (8) and (14) it follows that
unitarity is satisfied in this case in which there is a single self-consistent
solution both when there is, and is not, a BB incident at z = z$_{1}$ at t =
-$\epsilon$,

In case 1 the approach in HDP is equivalent to the imposition of the EKT
consistency principle; Eq. (7) has the effect of picking out the
self-consistent states $\uparrow\uparrow$ or $\downarrow\downarrow$ at t =
+$\epsilon$, as the only states which contribute to the path integral; this is
in accordance with the consistency principle, according to which only such
states are physical. The contribution of the states $\uparrow\downarrow$or
$\downarrow\uparrow$ , which do not satisfy the consistency condition and
would be regarded as unphysical by EKT, is suppressed by Eq. (7).

One can see in a simple way the connection in the HDP approach between the
preservation of unitarity and the existence of a consistent solution. In case
1, Eq. (7) for the X$_{ij}$ is equivalent to the statement that the operator X
can be written as X = U$_{\epsilon}$(T)U(T)U$_{\epsilon}$(0). Here
U$_{\epsilon}$(0) = U($\epsilon$, -$\epsilon$) is a unitary operator which
takes the initial states $\uparrow\downarrow$ and $\downarrow\downarrow$ at t
= -$\epsilon$ into the consistent states $\uparrow\uparrow$ and $\downarrow
\downarrow$ , respectively at t = $\epsilon$. Note that, for 0
$<$
t
$<$
T, we need consider only the two dimensional subspace, spanned by the
consistent states, of the full 4 dimensional Hilbert space, since the two
totally inconsistent states make no contribution to the operator X as given by
Eq. (7).\ Similarly, U$_{\epsilon}$(T) = U(T - $\epsilon$, T+ $\epsilon$)
takes $\uparrow\uparrow$ and $\downarrow\downarrow$ into the final states
$\uparrow\downarrow$ and $\downarrow\downarrow$,respectively, at t = T +
$\epsilon$; the relation U$_{\epsilon}$(T)$\uparrow\uparrow$ = $\uparrow
\downarrow$ reflects the disappearance of the BB at z$_{2}$ into the wormhole
a t = T. U(T) is the unitary time evolution operator from $\epsilon$ to T --
$\epsilon$ for the consistent states, given by Eqs. (9) and (10); The
appearance of the full operator U(T) in X is a consequence of consistency;
since U(T) leaves the wave function of a consistent state unchanged at z =
z$_{2}$, limiting the right side of Eq. (7) to terms diagonal in k imposes no
restriction. X is thus unitary since it is the product of three unitary
operators. While our argument is specific to the highly simplified HDP
approach, it seems likely that the conclusion that the unitaruty condition,
Eq. (8), holds will be valid in general when a unique consistent solution
exists in both the situation when there is, and when there is not, an incidnt BB.

\begin{center}
Case 2
\end{center}

Contrast case 1 with that in which we modify the operator U by replacing Eq.
(10) by U(T)$\uparrow\uparrow$= a$\uparrow\uparrow$ + b$\uparrow\downarrow,$
with
$\vert$%
a%
$\vert$%
$^{2}$ +
$\vert$%
b%
$\vert$%
$^{2}$ = 1; Eq. (11) remains unchanged. We can think of this U as simulating
the situation where there is a probability
$\vert$%
b%
$\vert$%
$^{2}$ that, upon emerging from the t = 0 mouth of the wormhole, the BB hits a
switch and transmits an electromagnetic signal closing a shutter and
preventing the incident BB from entering the wormhole at z = z$_{2}$ and t =
T. Eq. (7) now yields the result that
$\vert$%
X(1)%
$\vert$%
$^{2}$=%
$\vert$%
a%
$\vert$%
$^{2}.$.From unitarity, Eq. (13) is replaced by U(T)$\downarrow\uparrow$ =
a$\uparrow\downarrow$ - b$\uparrow\uparrow,$ which yields
$\vert$%
X(0)%
$\vert$%
$^{2}$ = 1+%
$\vert$%
b%
$\vert$%
$^{2}.$ Hence, unless b = 0, which is the self-consistent case 1, Eq. (8) does
not hold and X is not unitary. Note that, for U(T) obeying Eq. (11), if b
$\neq$ 0 there is no linear combination of $\uparrow\uparrow$ and
$\uparrow\downarrow$ on which the action of U(t) is given by that of an
operator with the direct product structure of Eq. (4). The violation of
unitarity is thus directly connected to the lack of a completely
self-consistent solution within the region containing CTC's, combined with the
consistency requirement that only matrix elements diagonal in k appear on the
right side of Eq. (7).

The billiard ball version of the grandfather paradox corresponds to the case
$\vert$%
b%
$\vert$%
$^{2}$ = 1 -
$\vert$%
a%
$\vert$%
$^{2}$ = 1, in which Eq. (10) is replaced by%

\begin{equation}
U(T)\uparrow\uparrow=\uparrow\downarrow\tag{10'}%
\end{equation}
ad we shall primarily concern ourselves with this case, in which the BB
emerges from the wormhole if and only if it does not enter it. We describe
this case as maximally inconsistent, since the overlap between U(T)$\Psi$(0)
and $\Psi$(0) at z = z$_{2}$ vanishes; this also maximizes the violation of
unitarity, since it gives
$\vert$%
$\Psi$(T + $\epsilon$)%
$\vert$%
$^{2}$ = 0. \ Note that this provides an explicit, albeit highly simplified,
example of a model in which no self-consistent solution exists, in contrast to
the situations discussed in Refs. 10 and 11.

We can interpret
$\vert$%
b%
$\vert$%
$^{2\text{ }}$as giving the probability that the apparatus producing the
interaction between the older and younger versions of the BB is functioning
correctly. \ A small nonzero value of
$\vert$%
a%
$\vert$%
$^{2}$ could arise, e. g., from a nonzero probability that the transmitter
which is to send the message closing the shutter and diverting the incident
ball from its original trajectory fails to send the message when the
transmitter is activated by a ball emerging from the wormhole at t = 0. Only
in the case of this, or similar, events, whose \textit{a priori} probability
one would expect in general to be very small, can the incident ball reach the
t = T mouth of the wormhole, thereby allowing the solution in which $\Psi
$(+$\epsilon$) = $\uparrow\uparrow$ to be self-consistent.

\bigskip One can seek to preserve the probability interpretation of the wave
function by renormalizing the final state $\Psi_{f}$ at t \
$>$
T by changing the prescription from $\Psi_{f}$ = X$\Psi$(-- $\epsilon$) \ to
$\Psi_{f}$ =NX$\Psi(-\epsilon)$, where N = [$\Psi^{\dag}(-\epsilon)X^{\dag
}X\Psi(-\epsilon)$]$^{-1/2}$; the normalizing constant N thus depends on the
initial state so that the time evolution is nonlinear. As discussed in HDP,
this implies that. if the initial state is a superposition of the states
$\uparrow_{1}$and $\downarrow_{1\text{,}}$the existence of the era of CTC's at
t
$>$
0 affects the time evolution of the system for t
$<$
0 in such a way that the probability of having n = 1 at t = 0 is proportional
to
$\vert$%
a%
$\vert$%
$^{2}$and thus vanishes in the maximally inconsistent limit a $\rightarrow0$.
\ Thus in the presence of a time machine we cannot impose initial conditions
at t = 0 so as to have the BB arriving at z = z$_{1\text{ }}$and a properly
functioning transmitter leading to Eq. (10') and the resulting potential
grandfather paradox.

Anderson[14\} suggests the alternative renormalization procedure of replacing
the evolution operator X by the unitary operator U$_{X\text{ }}=$ (X$^{\dag}%
$X)$^{-1/2}$X which is independent of the initial state. The two procedures
have the same effect when n = 1 at \ t = 0; \ in that case either procedure
results in multiplying the final state vector at t
$>$
T by 1/a. \ However, if one begins with a superposition of states with n = 1
and n = 0 the effects of the two renormalization procedures differ. \ In the
Anderson procedure.the time evolution for t
$<$
0 is unaffected by the presence of the time machine in the future. \ However,
in our simple model the evolution operator U$_{X}$ within the region of CTC's,
0
$<$
t
$<$
T is simply the identity operator. Thus all dependence on the physical
parameter a is eliminated by this procedure, and the actual evolution is the
same for an arbitrary value of a as for the case
$\vert$%
a%
$\vert$
= 1, in which case the transmitter is certain to fail and no inconsistency
arises. \ This procedure seems to lack a compelling physical motivation, and
to discard much of the essential physics associated with the CTC's. \ Both of
the renormalization procedures have the effect of rendering, essentially by
fiat, \textit{a priori} unlikely events certain. \ If one renormalizes the
final state, this happens at t
$<$
0, before the era of CTC's. \ In the procedure of Ref. 14, it occurs during
that era.

Both the renormalization prescriptions becomes undefined when a $=$ 0, i. e.
when the "grandfather paradox" is complete. This is, however, presumably an
unphysical limit; one expects that, in any actual case, there will always be
some nonzero probability, however small, for events to occur which would allow
the paradox to be evaded, so this objection is not conclusive. \ If there are
consistent classical solutions, these will dominate the path integral and only
the occupation numbers need to be treated quantum mechanically. \ When this is
true, one can expect that the path integral in the HDP model provides a
reasonable description of the actual behavior of the system. In the limit a
$\twoheadrightarrow$ 0 where there are no consistent classical solutions to
dominate the actual path integral, it will presumably be given by an integral
over many paths, and the argument that it is well represented by the path
integral evaluated in HDP ceases to be persuasive; the many degrees of freedom
of the BB-detector system not taken into account in HDP presumably become
important \ However, it seems very plausible that the qualitative conclusion
of the HDP model, namely, that the path integral for the case when there is an
incident BB at t = 0 is very small, will remain valid in this limit. \ One
expects this to be true precisely because of the absence of consistent
classical solutions, which would normally make the dominant contribution to
the path integral for a macroscopic system. \ Since on physical grounds one
does expect that consistent solution(s) will exist when n = 0 initially and
there is no incident BB, i. e., for the vacuum-to-vacuum process, we expect
that the conclusion that
$\vert$%
X(0)%
$\vert$%
$^{2}$
$>$
$\vert$%
X(1)%
$\vert$%
$^{2}$, and that hence, from Eq. (8), unitarity is violated when paradoxes
occur, will remain true in the full theory. The simple HDP model clearly
cannot be expected to give any detailed information about what actually occurs
in the a $\rightarrow$ 0 limit; \ that is, it will give no information about
the \textit{relative }probability of various \textit{a priori} improbable
events. But one does expect that the conclusion that seemingly improbable
events of some must occur if unitarity is to be preserved will remain valid.

One thus has, in the quantum mechanics of pure states, two rather unappealing
alternatives when there are no consistent classical solutions. First, one can
accept a nonunitary time evolution operator and the consequent loss of the
probability interpretation of quantum mechanics. Alternatively,, one can adopt
one of the renormalization procedures, meaning seemingly unlikely events
become certain either before or during the era of CTC's. This leads to what we
have previously referred to as the "banana peel" mechanism; while highly
counterintuitive, it at least allows the theory to be interpreted.

\begin{center}
Case 3
\end{center}

As a final example, we examine briefly the case in which there are two
self-consistent solutions. Take U(T) to be such that U(T) $\uparrow\downarrow$
= $\uparrow\downarrow$and U(T) $\uparrow\uparrow$ = $\downarrow\uparrow$. The
unitarity of U plus the physical requirement that U(T)$\downarrow\downarrow$ =
$\downarrow\downarrow$ then requires that U(T) $\downarrow\uparrow$=
$\uparrow\uparrow$. The states $\uparrow\uparrow$and $\uparrow\downarrow$ at t
= +$\epsilon$ now each give a self-consistent solution, and it is
straightforward to show from Eqs. (7) and (8) that unitarity is obeyed.
However, a nontrivial linear combination of the states $\uparrow\uparrow$ and
$\uparrow\downarrow$at t = $\epsilon$ does not result in a final state at T --
$\epsilon$ with the product structure of Eq. (5). \ Instead the state of the
system at t = T -- $\epsilon$ is of the correlated form c$_{1}\uparrow
\downarrow$ + c$_{2}\downarrow\uparrow$ which, from the discussion in Section
II, cannot represent a self-consistent pure state. The procedure in HDP, in
which the two consistent states $\uparrow\uparrow$ and $\uparrow\downarrow$
are included with equal weight in the path integral, thus implies that the
system is in a mixed state in the region containing \ CTC's. However, the two
solutions lead to orthogonal final states at t
$>$
T, and hence do not interfere with one another, so their undetermined relative
phase is irrelevant in Eq. (7).

Thus, in this example, Eq. (7) provides a unique and well behaved solution for
t
$>$
T, but within the region 0
$<$
t
$<$
T which contains CTC's one has a mixed state with two different
self-consistent solutuions which are present with equal probability. These two
solutions are physically quite different. This can be seen most clearly for
the initial condition in which there is no BB present at z$_{1}$ for t
$<$
0. Then for one of the two consistent solutions the system is in the state
$\downarrow\downarrow$ throughout the interval 0
$<$
t
$<$
T; this is precisely the solution one would naively expect. However, there is
also a consistent solution in which the system goes from the state
$\downarrow\uparrow$ to $\uparrow\uparrow$ in this interval. Here this second
solution can be dismissed on grounds of conservation of energy (i. e., of BB
number), but one does not expect this to be true in more realistic models
Intuitively, the first of these solutions seems more likely to be physically
relevant. Similar considerations apply to the case where n = 1 at t =
-$\epsilon$. Given the physical differences between the consistent
solutions,\ one might hope that, if CTC's are possible, a complete theory
would settle the uniqueness question, perhaps on the basis of entropy
considerations [12. 16, 17], and would select only one of the possible
consistent solutions in the interval 0
$<$
t
$<$
T. Then if one included only the physical solution in the path integral, case
3 would become mathematically identical to case 1; the system would be in a
pure state for all t, with a unique wave function obeying unitarity and having
the required continuity across the wormhole.

The form of U(t) for case 3 which we have been discussing appears somewhat
unphysical. in that involves nonconservation of BB number. This can be
avoided, while having two consistent solutions, only if U(T) is diagonal.
(This is an artifact of the simplicity of our model in which U is a 4x4
matrix, and there are only two possible states of the system for t
$>$
T) The case in which U is diagonal has an interesting property, in that it can
lead to the violation of unitarity even though consistent classical solutions
exist. \ This occurs because either in the case n = 1 or n = 0 at t
$<$
0, there are now two consistent solutions which contribute to the same final
state and, from Eq. (7), they will interfere with one another. \ If one
chooses U(t) such that the relative phase between the two possible state at t
= T - $\epsilon$ is different for the cases n = 0 and n = 1, the unitarity
condition, Eq. (8), will be violated. \ (The unitarity violation in the
example discussed in HDP in fact occurs for this reason, rather than because
of the nonexistence of classical solutions,) \ Once again, in our example this
phenomenon appears as an artifact of the overly simplified model. \ In
general, one would expect the states with n = 0 and 1 at t = T - $\epsilon$ to
lead to different final states at t
$>$
T so they would not give rise to quantum interference. \ The interference
occurs in the HDP model \ since there is only a single state containing (and a
single state not containing) a BB at t
$>$
T. \ In contrast, the type of unitarity violation found in case 1 would appear
to be generic in such situations, since it occurs because the \ magnitudes of
the individual terms on the right side of Eq. (7) become small compared to 1,
and does not depend on quantum interference.

\begin{center}
IV. Consistency, the Density Matrix, and the Mixed State MWI
\end{center}

As discussed in the introduction, the approach in DD requires, in general,
describing systems in the presence of CTC's in terms of their density matrix
$\varrho$ rather than a wave function. We begin this section by reviewing
briefly some simple ideas concerning the density matrix in the present
context. Next we show the connection that follows, in Deutsch's approach,
between the existence of potential \textquotedblleft grandfather
paradoxes\textquotedblright\ and the necessity for adopting a density matrix
description in regions containing CTC's. We then review the argument given in
DD that, in quantum mechanics with the MWI, CTC's do not lead to logical contradictions.

In the present case, in which there are four possible states of the system at
a given time which can be labeled by pairs of indices ij, where i,j = 0,1 give
the occupation numbers at z = z$_{1}$, z$_{2}$, the density matrix $\varrho$
is a 4x4 matrix whose matrix elements can be labeled $\varrho_{ij;mk}$. The
diagonal elements $\varrho_{ij;ij}$ give the probability that the system is in
the state with occupation numbers i and j; thus tr $\varrho$ = 1. For a pure
state, described by a wave function, $\varrho$ satisfies the condition
$\varrho^{2}$ = $\varrho$. The matrix elements $\hat{\varrho}_{2jk}$ of the
effective density matrix $\hat{\varrho}_{2}$ for the system at z = z$_{2}$ are%

\begin{equation}
\hat{\varrho}_{2jk}=\varrho_{ij;ik} \tag{15}%
\end{equation}
where the repeated index i Is summed over. The fact that the full system is in
a pure state does not imply that the density matrices $\hat{\varrho}_{i}$ for
the two subsystems satisfy the pure state condition. However, when the overall
state vector has the structure of a direct product, as in Eq. (16) below, the
full density matrix $\varrho$ also has a direct product structure $\varrho$ =
$\hat{\varrho}_{1}\otimes\hat{\varrho}_{2}$; the pure state condition on
$\varrho$ then implies that the separate systems at z = z$_{1}$ and z$_{2}$
are also pure states.

Let us return to our example of the BB which travels backward in time and
interacts with its younger self. Suppose we have a pure state at t =
+$\epsilon$. Since the occupation number at z = z$_{1}$ is 1 from the initial
conditions, the most general form of the state vector at is

\qquad\qquad%
\begin{equation}
\Psi(+\epsilon)=\Psi_{1}(+\epsilon)\otimes\Psi_{2}(+\epsilon)=\uparrow
\otimes(c_{1}\uparrow+c_{2}\downarrow) \tag{16}%
\end{equation}
with
$\vert$%
c$_{1}$%
$\vert$%
$^{2}$ +
$\vert$%
c$_{2}$%
$\vert$%
$^{2}$ = 1. The most general state of the system is one in which the system at
z$_{2}$, emerging from the wormhole, is in the pure state $\Psi_{2}$ =
(c$_{1}\uparrow$ +c$_{2}\downarrow$ ) with a definite phase relation between
the occupied and unoccupied components of the state vector. Continuity across
the wormhole then means that the state vector at t = T must have the form

\qquad%
\begin{equation}
\Psi(T)=U(T)\Psi(+\epsilon)=\Psi_{1}(T)\otimes\Psi_{2}(+\epsilon
)=(d\uparrow+d_{2}\downarrow)\otimes\Psi_{2}(+\epsilon) \tag{16'}%
\end{equation}
In particular, there must be no correlation between the occupation numbers at
z$_{1}$ and z$_{2}$. Thus for a given U(T), the consistency constraint can be
satisfied only if it is possible to choose the constants c$_{1}$ and c$_{2}$
so that $\Psi$(T) has the form given in Eq. (16'). In case 1 of Section III,
in which U(T) is given by Eqs. (10) and (11), Eq. (16') is obeyed, with
d$_{2}$ = 0, if we take the self-consistent solution with c$_{2}$ = 0. Of
course we already know this is a self-consistent pure state solution. since we
found that the time evolution operator X given by Eq. (7), which has the
consistency constraint built in, is unitary.

Now consider case 2 in which U(T) is given by Eqs. (10') and (11). There is
then no consistent solution and the operator X exhibits maximal violation of
unitary. Suppose we choose c$_{1}$ = c$_{2}$ = 1/$\sqrt{2}$, with both taken
to be real for simplicity; then, at t = +$\epsilon$, we have the product state,

\qquad\qquad\qquad%
\begin{equation}
\Psi(+\epsilon)=\uparrow_{1}(\uparrow_{2}+\downarrow_{2})/\sqrt{2} \tag{17}%
\end{equation}
in which the subsystems at z$_{1}$and z$_{2}$, and thus the full system are
all in pure states.

From Eqs. (17), (10'), and (11) one obtains\qquad\qquad\qquad\qquad
\qquad\qquad\qquad%

\begin{equation}
\Psi(T)=(\uparrow_{1}\downarrow_{2}+\downarrow_{1}\uparrow_{2})/\sqrt{2}
\tag{18}%
\end{equation}
This is not of the form of Eq. (16') because of the correlation between the
states at z$_{1}$ and z$_{2}$. Eq. (18) describes a situation in which the
subsystem at z = z$_{2}$, considered in isolation, is in a mixed state, with
no definite phase relation between the occupied and unoccupied states at z =
z$_{2}$ since the coefficients of $\uparrow_{2}$ and $\downarrow_{2}$ depend
on the coordinates of the system at z = z$_{1}$. (Recall that the symbol
$\uparrow_{1}$ is really shorthand in our case for the full wave function of a
BB, including the dependence on the internal coordinates.) We can also see
that the z$_{2}$-subsystem in Eq. (18) is not in a pure state by constructing
the density matrix $\varrho$ for the complete system and using Eq. (15) to obtain

\qquad%
\begin{equation}
\hat{\varrho}_{2}(T)=I_{2}/2 \tag{19}%
\end{equation}
where $\hat{\varrho}_{2}$(T) is the density matrix for the subsystem at z =
z$_{2}$ at t = T and I$_{2}$ is the 2x2 identity matrix. Eq. (19) describes a
mixed state with equal probabilities of
${\frac12}$
for finding a BB entering, or not entering, the wormhole at t = T. Since the
subsystem at z = z$_{2}$ is in a mixed state at t = T, this subsystem, and
therefore, from Eq. (17). the system as a whole, must be in a mixed state at t
= 0 if there is to be continuity across the wormhole. Thus, in the presence of
potential inconsistent causal loops, there are only two possibilities: the
first is that the continuity condition across the wormhole may not be exactly
satisfied, and when we project out the (possibly nonexistent) part of the wave
function satisfying the consistency condition the operator X becomes
nonunitary, as in case 2 in Section III; the second possibility is that pure
states are transformed into mixed states in the region containing CTC's.

It is demonstrated in DD that for any U(T) one can always choose values of
c$_{1}$ and c$_{2}$ in Eq. (16) that yield a $\varrho$ which, when evolved
according to Eq. (1), satisfies a modified consistency requirement of the form

\qquad\qquad\qquad%
\begin{equation}
\hat{\varrho}_{2}(T)=\hat{\varrho}_{2}(0) \tag{20}%
\end{equation}
Consistency in this sense is possible because one is working with $\varrho$
instead of a wave function. The consistency condition on $\varrho$ is
satisfied if the correlation is such that the probability for a BB to enter
the wormhole at t= T is the same as for one to emerge at t = 0. In the MSMWI
picture, a measurement of whether or not a BB emerges from the wormhole at t =
0 causes a branching into two MWI \textquotedblleft worlds\textquotedblright,
both of which remain components of the state of the system at t
$>$
0, and in one of which a BB enters the wormhole mouth at t = T. The
consistency condition on $\varrho$, but not on $\psi$, will then be satisfied
if the probabilities of a BB entering the wormhole at t = T, and of it's
emerging at t = 0, are equal even if a BB traveling backwards in time does not
emerge in the same \textquotedblleft world\textquotedblright\ which it left.
For example, while the wave functions in Eqs. (17) and (18) do not satisfy the
continuity condition across the wormhole, they each yield a density matrix
$\hat{\varrho}_{2}$ given by (19) so that Eq. (20) is satisfied.

\begin{center}
V. The MSMWI and Time Ttavel Paradoxes Involving Microscopic Objects
\end{center}

\qquad Before considering the case of a macroscopic object such as a billiard
ball or space ship, we examine how the MSMWI works in a situation where,
instead of a macroscopic object, we have a single electron. This is closely
analogous to the situation considered in DD. In the spirit of the MWI we
include as part of our system a measuring device, e g., an array of Cerenkov
counters surrounding the wormhole mouth, which can presumably be designed to
detect with arbitrarily high certainty whether or not an electron emerges from
the wormhole. We again consider the \textquotedblleft grandfather
paradox\textquotedblright\ situation with U given by Eqs. (10') and (11). If
an electron is detected, we can imagine the measuring device causes the
incident electron at z$_{1}$ to be deflected, e. g., by temporarily turning on
an electric field near z$_{1}$, so that it never reaches the wormhole.

Let the state of the device be designated by q, which will become one of our
dynamical variables along with the occupation numbers n and n' at z$_{1}$ and
z$_{2}$. The matrix elements of the full density matrix $\varrho$ appearing in
Eq. (15) will now be labeled by two sets of three indices, plus t. Since we
are dealing with mixed states with undefined relative phases, the density
matrices are diagonal and we can specify the \ nontrivial density matrix
elements at time t uniquely by a single set of three indices, writing the
matrix elements as $\varrho$(q, n, n', t). (The matrix elements of the
effective density matrix $\hat{\varrho}$ for the system at z$_{2}$ will still
be labeled only by the values n', however; q and n are both degrees of freedom
associated with the remainder of the system and are both summed over in
finding $\hat{\varrho}$.) The Hamiltonian H and thus U in Eq. (1) will now
include H$_{m}$, the interaction between the electron and the measuring
device. Initially, at t
$<$
0, we take q = q$_{A}$, while q = q$_{B}$ after the detection of an electron
by the device. Thus, if an electron emerges from the wormhole at t = 0, then
at t = $\delta$
$>$
$\epsilon$, q becomes equal to q$_{B}$; $\delta$ is a property of the
detection system, and will be finite (though we assume $\delta$
$<$%
$<$
T), both because the counters will have a finite response time and because
they will be located at a finite distance from the position of the electron at
t = 0, the earliest time at which it could be observed as it emerges from the wormhole.

According to the picture in DD, for t
$>$
0 the system will be in a mixture of two states, each with probability
${\frac12}$%
. We will label these states A and B according to the values, q$_{A}$ and
q$_{B}$, respectively, of q at t = T -- $\epsilon$. Since q = q$_{A}$ in state
A, in that state no electron was detected at t = $\delta$, and it then follows
from Eqs. (10') and (11) that, in state A, n ` = 1 at t = T -- $\epsilon$ and
the incident electron enters the wormhole mouth at t = T. Similarly, in state
B, with q = q$_{B}$ at T -- $\epsilon$, no electron enters the wormhole at t = T.

Thus, at t = $\epsilon$
$<$
$\delta$, one will have a system with n = 1 and q = q$_{A}$ in a mixed state
with equal probabilities for finding n' = 0 and n' = 1. At t = $\delta$, q
becomes equal q$_{B}$ in the state with n' = 1; that is, the electron that was
in state A at t = T -- $\epsilon$ emerges from the wormhole at t = 0 and is
detected in state B at t = $\delta$.

For t
$>$
$\delta$ an observer, as in the conventional MWI, has an equal chance of being
in the \textquotedblleft worlds\textquotedblright\ with q = q$_{A}$ or
q$_{B\text{ }}$. \ In state B, with q = q$_{B}$, the observer sees the
electron initially at z$_{1}$ deflected so that it never reaches the wormhole,
while the electron leaving the wormhole arrives at z$_{1}$ at t = T, in
accordance with Eq. (10'), so that n'= 0 at t = T, and the observer will
conclude that n'(0) $\neq$ n'(T). A similar analysis holds for observers in
the \textquotedblleft\ world\textquotedblright\ in which q = q$_{A}$ for t
$>$
$\delta$. The time evolution during the period 0
$<$
t
$<$
T will appear perfectly sensible to observers in both worlds. They will be
surprised to see that n'(T) $\neq$ n'(0), but this does not constitute an
actual logical contradiction, since n'(0) and n'(T) are physically different
observables for outside observers so that the theory does not give
contradictory predictions for the value of the same observable as seen by the
same observer.

A hypothetical observer riding on the electron will also see nothing unusual.
The electron apparently evolves normally, in terms of the local time variable
$\tau$ and Hamiltonian H' discussed in Section II, in passing through the
wormhole; an observer moving with the electron would see n'($\tau$ = T) =
n'($\tau$ = T + t$_{0}$) = 1, where t$_{0}$ is the transit time through the
wormhole, and will see outside clocks reading t = 0 as he emerges. However,
the world in which he now finds himself will be different than he saw when
$\tau$ = t =0 and the electron was at z = z$_{1}$, since now he will see q =
q$_{B}$, and find himself in a world with two electrons.

Thus the communication between different MWI \textquotedblleft
worlds\textquotedblright\ postulated in Deutsch\textquotedblright s approach
actually occurs, if the continuity condition is given by Eq. (20), as a result
of the interaction H$_{m}$ with the measuring apparatus. The electron enters
the wormhole in the q = q$_{A}$ \textquotedblleft world\textquotedblright\ and
appears at t = 0, also in a state with q = q$_{A}$.. However, at t = $\delta$,
when the measurement process results in branching into two separate
\textquotedblleft worlds\textquotedblright\ with different values of q, q
becomes equal to q$_{B}$ in the state containing the electron at z$_{2}$. The
electron, which is in state A at t = T, is able to appear in state B at a
later value of its own time $\tau$ by traveling back in time through the
wormhole to a time t
$<$
$\delta$ before the measurement, and the resulting branching into states A and
B, has occurred.

More formally, we can understand this as follows. Let us consider the time
translation operator U(t$_{2}$, t$_{1}$) $\equiv$U$_{21}$ for the case t$_{2}$
= $\delta$ + $\epsilon$', t$_{1}$ = T, with $\epsilon$' arbitrarily small and
$\delta$
$<$%
$<$
T. We can write U$_{21}$ = U($\delta$+$\epsilon$',0)U(0,T) = U($\delta$)U(-T);
U(-T) is the analog of the corresponding operator in Eq. (4), but does not
have a direct product structure because of the correlations between
observables at t = T. At t = T the system will be, with equal probability, in
states with q =q$_{A}$, n' = 1 and q = q$_{B}$, n' = 0, so the nonzero
diagonal density matrix elements $\varrho(q,n,n^{\prime},t)$ for t = T will be

\qquad%
\begin{equation}
\varrho(q_{A},n(T)_{A},1,T)=%
\frac12
=\hat{\varrho}_{2}(1,T) \tag{21a}%
\end{equation}
\qquad and

\qquad\qquad\qquad\qquad%
\begin{equation}
\varrho(q_{B},n(T)_{B},0,T)=%
\frac12
=\hat{\varrho}_{2}(0,T) \tag{21b}%
\end{equation}
where n(T)$_{A}$, e. g., is the value of the occupation number n at t= T and q
= q$_{A}$. T he final equality in Eqs. (21) is a consequence of the fact that
only the matrix elements of $\varrho$ appearing in Eqs. (21) are nonzero.

The operator U(-T) transforms $\varrho$(T) into $\varrho$(0). By Eq. (20),
this must leave $\hat{\varrho}_{2}$ invariant, while at t = 0 the only nonzero
elements of $\varrho$ are for n = 1 and q =q$_{A}$. Thus at t = $\epsilon$
$<$
$\delta$ the nonzero elements of $\varrho$ are%

\begin{equation}
\varrho(q_{A},1,n\prime,\epsilon)=\hat{\varrho}_{2}(n\prime,\epsilon)=%
\frac12
,\text{ \ }n\prime=0,1 \tag{22}%
\end{equation}
Comparing Eqs. (21) and (22), we see that Eq. (20) is indeed satisfied.

Since $\delta$
$<$%
$<$
T, the factor U($\delta$) in U$_{21}$ differs from unity only because of the
interaction H$_{m}$ with the measurement device. Thus, acting on states
$\vert$%
q,n,n'%
$>$%
, U($\delta$)%
$\vert$%
q$_{A}$,1,1%
$>$
=
$\vert$%
q$_{B}$,1,1%
$>$%
, while U($\delta$) leaves
$\vert$%
q$_{A}$, 1, 0%
$>$
unaffected. Thus%

\begin{equation}
U_{21}|q_{A},n(T)_{A},1>=|q_{B},1,1> \tag{23a}%
\end{equation}
\qquad and%

\begin{equation}
U_{21}|q_{B},n(T)_{B},0>=|q_{A},1,0> \tag{23b}%
\end{equation}
From Eq. (1), there will be a similar transformation of the diagonal density
matrix elements $\varrho$(q, n, n', t) so that, for n' = 1, we have from Eqs.
(21a), (22), and (23a)

\qquad\qquad\qquad%
\begin{equation}
\varrho(q_{A},n(T)_{A},1,T)=\varrho(q_{A},1,1,0)=\varrho(q_{B},1,1,\delta)=%
\frac12
\tag{24a}%
\end{equation}
and, similarly, for n' = 0%

\begin{equation}
\varrho(q_{B},n(T)_{B},0,T)=\varrho(q_{A},1,0,0)=\varrho(q_{A,}1,0,\delta)=%
\frac12
\tag{24b}%
\end{equation}
and one sees that the electron, which entered the wormhole at t = T in state
A, is found at t = $\delta$ in state B. The continuity condition (20) on the
subdensity matrix 2 is satisfied, since the probabilities for finding n' = 0
and of finding n' = 1 are both equal to one-half at each end of the wormhole.

Thus the MSMWI, with an object described by a density matrix satisfying Eq.
(20), leads, as asserted in DD, to a quantum theory of a microscopic object
passing through a time machine which avoids the \textquotedblleft grandfather
paradox\textquotedblright. This occurs, as in the parallel universes of
science fiction, because the object emerges from the time machine and
\textquotedblleft murders\textquotedblright\ its younger self in a different
\textquotedblleft world\textquotedblright, i. e., an orthogonal \ quantum
state, when it travels back in time.

\begin{center}
VI. The MSMWI for Macroscopic Objects
\end{center}

As we now show, however, problems arise if one applies the MSMWI in the case
of macroscopic objects, such as billiard balls, passing through the wormhole.
We first specify the meaning we will attach to macroscopic in this context.
Let the object in question have linear dimension d in its direction of motion
and be moving with speed v, so that it requires a time interval $\Delta$t =
d/v to emerge from the wormhole. That is, for 0
$<$
t
$<$
$\Delta$t, the front portion of the BB exists on a timelike surface t =t$_{1}$
while the back portion exists on the timelike surface t = t$_{1}$ + T. We will
call the object macroscopic if $\Delta$t
$>$
$\delta$, where $\delta$ is the time at which the detector recognizes that the
object has emerged, and in consequence sends a signal preventing the object
from entering the wormhole at t = T; as in Section V, $\delta$ depends on the
resolution time of the detector and its distance from the position of the
leading edge of the object as it emerges at t = 0. Since a fraction f =
$\delta$/$\Delta$t of the object must emerge from the wormhole before the
detector is triggered, for a macroscopic object f
$<$
1 and a fraction 1 -- f
$>$
0 of the object will not yet have emerged from the wormhole at t = $\delta$

The above definition of "macroscopic" has the problem of depending on $\delta
$, and thus on the particular detection device being used. \ One can introduce
a more fundamental definition to avoid this by taking $\Delta t>\delta_{\min}%
$, where $\delta_{\min}\approx$ 1/m, with m the mass of the object and c =
1,is the smallest possible resolution time for any detector.\ Then on the
fundamental level we would take an object to be macroscopic if \ d
$>$
1/m or md
$>$%
1.

The HDP model must be extended somewhat to accommodate macroscopic objects,
but the generalizations do not change the physics in an essential way. Clearly
the wormhole mouth must have a finite radius. Also, the wormhole must persist
for a time T$_{0}$
$>$
$\Delta$t in order for the object to traverse it. One must then generalize the
wormhole to identify times t and t + T for 0
$<$
t
$<$
T$_{0}$. where $\Delta$t
$<$
T$_{0}$
$<$
T; the upper limit on T$_{0}$ avoids the necessity of introducing a spatial
separation between ends of the wormhole which overlap in time, as discussed
earlier. Eq. (20) must be correspondingly generalized to \qquad\qquad

\qquad\qquad\qquad%
\begin{equation}
\hat{\varrho}_{2}(t+T)=\hat{\varrho}_{2}(t) \tag{20'}%
\end{equation}
\qquad\qquad\qquad\qquad

We will place one additional restriction on the wormhole persistence time
T$_{o}$. \ Let T$_{s}$ be the time at which the incident BB reaches the
shutter whose closure prevents it from entering the wormhole. \ We will
strengthen the restriction on T$_{o}$ by requiring T$_{o}$
$<$%
\thinspace\ T$_{s}$. We thus eliminate the possible consistent solution
mentioned in the Introduction, in which the BB squeezes past the shutter just
as it closes, being slowed down in the process so that it reaches the wormhole
at t =T + T$_{s}$, and reemerges at t = T$_{s}$ to trigger the shutter just as
its younger self reaches it. \ According to the EKT consistency principle,
this would become the physically observed process, thus evading the paradox.
However, this consistent solution does not exist if the early-time mouth of
the wormhole closes before the BB reaches the shutter, thus eliminating the
possibility of a BB emerging from the wormhole at t = T$_{s}$ and triggering
the shutter just as the incident BB reaches it.

Let us consider first, for simplicity, the case f =
${\frac12}$%
. + $\epsilon$'; i. e., we assume that, on the average, just over half of the
BB emerges from the wormhole before the detector is triggered. By analogy with
our discussion in the previous section, at t = T -- $\epsilon$, in state A,
with q= q$_{A}$, the incident BB will be about to enter the wormhole, since in
that state the detector was not triggered at t = $\delta$, while in state B
there will be no BB entering the wormhole. Then at t
$<$
$\delta$ one will have a mixture of two states, both with q = q$_{A}$, in one
of which the front portion of a BB will have emerged from the wormhole; this
latter state will be state B, with q = q$_{B}$ for t
$>$
$\delta$, since in this state the detection device will be triggered. \ The
density matrix ar t = $\epsilon$
$<$
$\delta$ will be given by Eq. (22), where n' = 1 denotes the presence of the
front edge of the BB at z$_{2}$

For t
$>$
$\delta,$ q is a constant of the motion since H$_{m}$, the interaction
Hamiltonian with the detection device, has no matrix elements between the
states with q = q$_{A}$- and q = q$_{B}$, after the irreversible measurement
has been completed. This is the exact analog of the independence of different
\textquotedblleft worlds\textquotedblright\ from one another in the
conventional MWI without CTC's.

This decoupling of states A and B has far reaching consequences for the
predicted behavior of a macroscopic object passing through a wormhole. As with
the electron, the front half of the BB, which is in state A at t = T, appears
in state B at t = $\delta$ and $\tau$ = T + $\delta$. This can occur because
the front half travels back in time to the range of times 0
$<$
t
$<$
$\delta$ at which time q = q$_{A}$ in both states A and B. However, the rear
half of the BB reaches the wormhole mouth at t = t$_{1}\equiv$ T + $\Delta$t/2
= T + $\delta,$and hence it begins to emerge from the wormhole at $\ $t =
$\delta$, after the measurement has occurred. For t
$>$
$\delta$ the evolution operator U'$_{21}$(t, t$_{1}$) = \ U(t, $\delta$,)
U($\delta,$ T + $\delta$), plays the analogous role for the back half of the
BB that U$_{21}$ played for the electron in Section V; U$_{21}^{\prime}$ does
not connect states A and B, since, for t
$>$
$\delta,$ q$_{B}$ $\neq$ q$_{A}$ and these states are decoherent. For t
$>$
$\delta$ Eq. (23a) must be replaced by

\qquad\qquad\qquad\qquad%
\begin{equation}
U_{21}^{\prime}(t)|q_{A},n(T)_{A},1>=|q_{A},1,1> \tag{25}%
\end{equation}
and hence, from Eq. (1), the analog of Eq. (24a) for the nonzero matrix
elements of the density matrix for n' = 1 at times t and T + t at opposite
ends of the wormhole, when t
$>$
$\delta,$ is

\qquad\qquad%
\begin{equation}
\varrho(q_{A},n(T)_{A},1,T+t)=\rho(q_{A},1,1,t)=%
\frac12
,t>\delta\tag{26}%
\end{equation}
with analogous changes occurring in Eqs. (23b) and (24b). Hence there is
vanishing probability of finding the back half of the BB at z = z$_{2}$ in the
\textquotedblleft world\textquotedblleft\ with q = q$_{B}$, and the back half
of the BB, \textit{unlike the front half}, will necessarily emerge from the
wormhole in state A with q = q$_{A}$.

The MSMWI thus predicts that the two halves of the BB will emerge from the
wormhole in different MSMWI worlds! An external observer will, with
probability one half, see nothing emerge from the wormhole during the interval
0
$<$
t
$<$
$\delta$, so that the detection device is not triggered, and will end up in
state A with q = q$_{A}$. This observer will then see the rear half of the BB
emerge between t = $\delta$ and t = 2$\delta$ and go off to reach z = z$_{1}$
in accordance with Eq. (10'). Since the detector was not triggered, the
\textquotedblleft younger\textquotedblright\ BB initially at z$_{1}$ at t = 0
will not be deflected and will enter the wormhole between t = T and t = T +
2$\delta$. The front half of the BB, which entered the wormhole at T
$<$
t
$<$
T + $\delta$, will seem to this observer to have disappeared, since it emerged
in the other \textquotedblleft world\textquotedblright; this is similar to the
microscopic case. The rear half of the BB will match the rear half which
emerged earlier at t = $\delta$, so that observations at the two wormhole
mouths at t and t + T will indicate continuity across the wormhole for t
$>$
$\delta$, once the discontinuous measurement process, has been completed.

There will also be probability one half of observing the front half of a BB
emerging from the wormhole between t = 0 and t = $\delta$, triggering the
detection device, putting q = q$_{B}$, and causing the deflection of the young
BB, which therefore never reaches the wormhole. In this q$_{B}$
\textquotedblleft world\textquotedblright, nothing enters the wormhole mouth
at t
$>$
T and the front half of the BB will seem to appear for no apparent reason;
this is again similar to the electron case. However, for t
$>$
$\delta$, in state B nothing enters the wormhole at t + T or emerges at t, so
that, as in state A, external observers will see continuity between the two
mouths of the wormhole for t
$>$
$\delta$, after the two \textquotedblleft worlds\textquotedblright\ have decoupled.

This surprising result is possible because the continuity condition, Eq.
(20'), which is the basic assumption in DD, only constrains the elements of
the effective density matrix $\hat{\varrho}_{2}$. The density matrix elements
of the macroscopic BB must now be labeled by separate occupation numbers
n$_{f}^{\prime}$ and n$_{b}^{\prime}$ for the front and back segment of the
BB. The continuity of $\hat{\varrho}_{2}$ ensures that the total probability
of finding n$_{f}^{\prime}$ = 1, i. e., of detecting the front segment at
z$_{2}$, is one-half at each mouth of the wormhole. However, the matrix
elements of $\hat{\varrho}_{2}$ for a given value of n$_{f}^{\prime}$ involve
the sum over q of the matrix elements of the full density matrix $\varrho$ for
that value of n$_{f}^{\prime}$, and thus the relation between the values of
n$_{f}^{\prime}$ and the value of q need not be preserved in going through the
wormhole. The same holds true for n$_{b}^{\prime}$. In fact, as we have
seen,where, classically, there is a \textquotedblleft grandfather
paradox\textquotedblright, the relation between q and n' develops a
discontinuity. At t = T, just before the BB enters the wormhole, the set of
observables (q, n$_{f}^{\prime}$, n$_{b}^{\prime}$) have, with equal
probability, the sets of values (q$_{A}$, 1 1) and (q$_{B}$, 0, 0). However,
for the emerging BB at t
$>$
$\delta$, the relation between q and n$_{f}^{\prime}$ and between q and
n$_{b}^{^{\prime}}$differs from that for t
$<$
$\delta$ because of the discontinuous change in the value of q resulting from
the measurement, and the possible sets of values become (q$_{A}$, 0, 1) or
(q$_{B}$, 1, 0). There are equal probabilities at each end of the wormhole of
finding each possible value, 0 or 1, for both n$_{f}^{\prime}$ and
n$_{b}^{\prime}$, as required by Eq. (20'). However, the correlation between
the values of n$_{f}^{\prime}$ and n$_{b}^{\prime}$ for a given value of q is
different at the two ends of the wormhole. At t = T an observer in an MWI
\textquotedblleft world\textquotedblright\ with a definite value of q sees
nonzero values of the density matrix elements for the same values of
n$_{f}^{\prime}$ and n$_{b}^{\prime}$; i. e., he sees either the whole object
or nothing entering the wormhole. At the other end such an observer sees
nonzero probabilities for different values of n$_{f}^{\prime}$ and
n$_{b}^{\prime}$ and thus observes only the front or back half of the object.
For an elememtary particle \ this problem does not arise since the concept of
different parts of such an object is meaningless; for such an object, the
discontinuity due to the measurement is simply that associated with the
emergence of the object, which occurs suddenly rather than over time as in the
macroscopic case.

\qquad We can generalize the above discussion to other values of the fraction
f. Suppose, e. g., that f = 1/3 + $\epsilon$', meaning the detection device
can detect the emergence of one third of a BB, and $\delta$ = $\Delta$t/3. Let
us also assume that the detector, after being triggered, reads q$_{B1}$ or
q$_{B2}$, respectively, depending on whether it was triggered at t = $\delta$
by observing the first third of an emerging BB, or at t = 2$\delta$ by the
middle third. In both of these \textquotedblleft worlds\textquotedblright,
since the detector was triggered, the incident BB will not enter the wormhole
at t = T. Since q is a constant in these \textquotedblleft
worlds\textquotedblright\ for t
$>$
$\delta$ or t
$>$
2$\delta$, respectively, they will not be coupled to the third
\textquotedblleft world\textquotedblright\ for t
$>$
2$\delta$, and in neither of them will the last third of the BB be observed.
In this third world, the detector will not be triggered so that q remains
equal to q$_{A}$. It will couple only to itself for t
$>$
2$\delta$, and hence in the q$_{A}$ world one will observe the rear third of
the BB emerging from the wormhole between t = 2$\delta$ and t=3$\delta$.

There will thus be three MSMWI \textquotedblleft worlds\textquotedblright. The
solution satisfying the consistency condition (20') on the density matrix is
that each of these occurs with probability 1/3. There is then a one third
probability of having q = q$_{A}$ and a BB entering the wormhole at t + T.
This leads to probability 1/3 for each segment to emerge in its respective
\textquotedblleft world\textquotedblright, so that Eq. (20') is, indeed, satisfied.

More generally, let f = 1/N, where N is arbitrary, thus including the case of
a detector of arbitrarily high sensitivity. One would then have N MSMWI
\textquotedblleft worlds\textquotedblright, in each of which a fraction 1/N of
the BB would be seen to emerge during a time interval (i -- 1) $\Delta$t
$<$
i$\Delta$t, i $\preceq$I N. As N becomes arbitrarily large and the detector
becomes very sensitive, the probability of observing the BB actually reaching
the wormhole at t = T without being deflected thus vanishes as 1/N; in this
limit one will observe, essentially with certainty, a microscopic fragment of
the BB, which might be indistinguishable from random background, emerging from
the wormhole at some time between t = 0 and t = $\Delta$t, triggering the
detector, and preventing the incident ball from entering the wormhole; thus in
the limit of large N the probability of seeing the incident BB enter the
wormhole becomes vanishingly small, but the number of fragments into which it
is split becomes very large, so that the probability of some fragment emerging
in any one of the essentially infinite number of \textquotedblleft
worlds\textquotedblright\ is unity.

The fact that fractions of the BB emerge from the wormhole in states with
different values of q means that, if the MSMWI is correct, the Hamiltonian H'
controlling the evolution of the BB in its proper time $\tau$ through the
wormhole cannot be anything like that of a free BB; it must include violent
interactions with the matter and/or gravitational fields of the wormhole which
lead to the disintegration of the BB. The effect of these interactions is
presumably independent of the sensitivity of the device used to detect the
emerging BB. Thus it would appear that the MSMWI implies that a macroscopic
object traversing a wormhole (or other time machine) must necessarily be
broken up into microscopic constituents, presumably elementary particles,
which will appear pointlike to the most sensitive detectors possible. This
would, e. g., be true in a theory is which stable wormholes can exist only if
their radii are of the order of the Planck length. \ Such a wormhole would not
be \textquotedblleft traversable\textquotedblright\ in the sense of Ref. [1].
Hence the MSMWI does not provide a quantum theory which is free of paradoxes
and which describes wormholes, or similar objects involving CTC's, which are
traversable by macroscopic objects. \ 

\begin{center}
VII. Conclusions
\end{center}

\qquad We have considered two general approaches to resolving the problem of
apparent paradoxes in theories with CTC's. The first, illustrated by the
simplified model presented in HDP, attempts to preserve the quantum mechanical
notion of pure states and imposes an appropriate continuity condition, Eq.
(7), across a wormhole or other time machine on the wave function. When the
time evolution operator is such that there is a single self-consistent
solution, Eq. (7) is equivalent to the EKT consistency principle, and leads to
a theory which is both consistent and unitary. If there are multiple
consistent solutions, all of which are physical, problems arise because the
solution is not uniquely specified by requiring consistency. Hopefully these
would be absent in a more complete theory, which solves the uniqueness problem
by providing a procedure for selecting only one of the consistent solutions as physical.

\qquad However, it is possible to choose the Hamiltonian, in the HDP model, so
that no self-consistent solution exists, thus simulating the existence of
initial conditions at t = 0 leading to the formation of inconsistent causal
loops as in the \textquotedblleft grandfather paradox\textquotedblright\ It
seems likely that this is also true in more realistic models. Then the attempt
to enforce consistency by means of the continuity constraint in Eq. (7) leads
to a violation of unitarity in the operator X connecting the state vectors
before and after the region of CTC's. The probability interpretation of
quantum mechanics can be preserved only by renormalizing the final state or
the operator X. The renormalization has the effect of forcing the probability
of some events, e. g., the failure of a piece of apparatus, which would
normally be very small, to become equal to one; depending on the
renormalization procedure, the events in question may occur prior to the
construction of the time machine, i. e., prior to the formation of a Cauchy
horizon. Thus, postulating the renormalization procedure required to conserve
probability amounts to postulating the \textquotedblleft banana
peel\textquotedblright\ mechanism, i. e., the certain occurrence of some
member of a set of \textit{a priori }improbable events which conspire to
prevent paradoxes from occurring. The renormalization process fails if the
norm of the final state is strictly zero, i. e., if X is singular, meaning
there is no sequence of events, however improbable \textit{a priori}, by which
the paradox can be evaded.

The alternative approach in DD involves attempting to implement the idea of
parallel universes from science fiction so that apparently contradictory
events occur in different \textquotedblleft worlds\textquotedblright; if
successful, this would preserve the freedom to impose initial conditions
arbitrarily. This approach involves two fundamental assumptions: 1) In the
presence of CTC's, the MWI as given in Ref. 13 is correct, and not simply an
interpretation of quantum mechanics which one is free to adopt or not
according to taste. 2) In the presence of CTC's, individual systems may not be
in pure states but in mixed states characterized by a density matrix but not a
wave function. This differs from Ref. 13, in which physical systems, with the
measuring apparatus included, are taken to be in pure states; we therefore
refer to this as the MSMWI, where MS stands for \textquotedblleft mixed
state\textquotedblright. Assumption 2) has two corollaries. First, the concept
of the density matrix is extended to apply to single systems, in contrast to
its usual application to ensembles of systems that have been identically
prepared. Secondly, the correct formulation of the continuity condition in the
presence of a wormhole is not, in general, as a condition on a wave function
in the form of Eq. (7), but rather as the condition (20') on the density matrix.

\qquad For the potentially paradoxical case in which the time evolution
operator appears to be such that an object emerges from the wormhole at t = 0
if and only if it does not enter the wormhole at t = T, the mechanism
suggested in DD for resolving the paradox can be successful if the object is
microscopic. The different \textquotedblleft worlds\textquotedblright\ of the
MSMWI correspond to states in which a macroscopic detector, which records
whether the object emerged from the wormhole, has different readings, and are
thus effectively decoupled. A microscopic object is able to appear intact at t
= 0 in a different world from that in which it entered the wormhole because it
emerges from the wormhole at t = 0 before the measurement leading to the
branching of the worlds, has occurred.

If the object is macroscopic, however, it emerges from the wormhole over a
finite period of time. If this is greater than the resolving time of the
detector, the measurement, and the branching into two or more decoherent
states, will occur before the object has emerged completely from the wormhole.
The MSMWI \textquotedblleft worlds\textquotedblright\ then become decoupled,
as in the conventional MWI, and the subsequent segment or segments of the
macroscopic object cannot emerge in the same \textquotedblleft
world\textquotedblright, i. e., in the quantum state with the same reading of
the macroscopic detector, as the leading segment. The object is thus split
into a number of pieces in its passage through the wormhole. A given observer,
who sees a particular reading of the macroscopic detection device, will see
only one of these pieces.\qquad

\qquad The mechanism for eluding the grandfather paradox, proposed in DD, thus
appears to imply that macroscopic objects, when traversing a wormhole, undergo
interactions which are sufficiently violent as to break up the object. The
number of pieces into which the object is observed to be broken depends on the
sensitivity of the detector, but becomes very large if the detector is
sensitive enough to detect very small fragments. One expects the interaction
between the object and the wormhole should be independent of the sensitivity
of the device used to detect the emerging object. Hence one concludes that
assumptions 1) and 2), together with their corollaries, can be valid only if
macroscopic objects passing through a time machine interact with it so
strongly as to be disintegrated into fragments which appear pointlike to the
most sensitive possible detectors. One obvious class of theories in which this
would be true is the class in which stable wormholes can exist only if their
dimensions are of the order of the Planck length.

The approach in DD, therefore, does not provide an explanation of how
paradoxical results can be evaded in a theory with traversable wormholes, or
other kinds of traversable CTC's, where \textquotedblleft
traversable\textquotedblright\ is used in the sense of Ref. [1] as meaning
traversable intact by macroscopic objects such as billiard balls, space ships,
or human beings. Hence the only satisfactory candidate for a theory of such
objects appears to be one in which the necessity of renormalizing the future
scattering matrix constrains physics in the present in such a way that
conditions whose a priori probability seems very small, e. g., the presence of
a banana peel, are in fact rendered certain. \ \ \ \ 

\begin{center}
Acknowledgments
\end{center}

I am grateful fo Larry Ford, Tom Roman, and Serge Winitski for useful
discussions, and for helpful comments and criticisms of the manuscript.

\begin{center}
References
\end{center}

[1] M. Morros, J. Thorne, and Y Yurtsever, Phys. Rev. Lett. \underline{61},
1446 (1988) \ 

[2[ M. Alcubierre, Claas. Quant. Grav. \underline{11}, 173 (1994);

S. Krasnikov, Phys. Ref. D \underline{57}, 4760 (1998)

[3] A. Everett, Phys Rev. D \underline{53}, 7365 (1996).

[4] A. Everett and T. Roman, Phys. Rev. D \underline{56}, 2100 (1997).

[5] A. Ori and Y Soen, Phys. Rev. D \underline{49} , 3990 (1994)/

[6] S. W. Hawking, Pjys. Rev. D \underline{46}, 603 (1993).

[7] L. Ford and T. Roman, Phys. Rev. D \underline{51}, 4277 (1996);
\textit{ibid}. 63, 5496 (1996).

[8] M. Pfenning and L. Ford, Class. Quant. Grav. \underline{14}, 1743 (1997).
See also Ref. 4.

[9] M. Cassidy, Class. Quant. Gtav. \underline{14}, 523 (1997);

J. R. Gott and L-X. Li, Phys. Rev. D \underline{58}, 023501 (1998).

M. Visser, S. Kar, and N. Dadhich, Phys. Rev. Lett. \underline{20}, 201102 (2003.

[10] F. Echevaria, G. Klinkhammer, and K. Thorne, Phys. Rev. D \underline{44},
1077 (1991).

[11] I Novikov, Phys. Rev. D \underline{45}, 1989 (1992).

[12] H. D. Politzer, Phys. Rev. D \underline{49}, 3981 (1994).

[13] J. Hartle, Phys, Rev. D \underline{49}, 6543 (1994).

[14]A. Anderson, Phys. Rev. D\underline{51}, 5707 (1995).

[15] H. Everett, III, Rev. Mod. Phys. \underline{29}, 454 (1957).

[16] D. Deutsch, Phys. Rev. D \underline{44}, 3197 (1991); Sci. Amer. March,
1994, p. 68.

\bigskip\lbrack17] G. Romero and D. Tores, Mod. Phys. Lett. A\underline{16},
(1213) 2001.

\bigskip
\end{document}